\documentclass[reprint, twocolumn, superscriptaddress, nofootinbib,showkeys, floatfix, longbibliography, aps, amsfonts,amsmath,amssymb]{revtex4-2}
\usepackage{hhline}
\usepackage{graphicx,psfrag,xcolor}
\usepackage{bm}
\usepackage{hyperref}
\usepackage{multirow}
\definecolor{linkcolor}{HTML}{0176ba}
\definecolor{urlcolor}{HTML}{0176ba} 
\definecolor{citecolor}{HTML}{900020}
\hypersetup{pdfstartview=FitH,  linkcolor=linkcolor,urlcolor=urlcolor, citecolor=citecolor, colorlinks=true}

\begin{document}
\author{Nikita A. Ustimenko}
\email{nikita.ustimenko@metalab.ifmo.ru}
\affiliation{School of Physics and Engineering, ITMO University, St. Petersburg 197101,  Russia}
\author{Kseniia V. Baryshnikova}
\email{k.baryshnikova@metalab.ifmo.ru}
\affiliation{School of Physics and Engineering, ITMO University, St. Petersburg 197101, Russia}
\author{Roman V. Melnikov}
\affiliation{School of Translational Information Technologies, ITMO University, St. Petersburg 197101, Russia}
\author{Danil F. Kornovan}
\affiliation{School of Physics and Engineering, ITMO University, St. Petersburg 197101, Russia}
\author{Vladimir I. Ulyantsev}
\affiliation{School of Translational Information Technologies, ITMO University, St. Petersburg 197101, Russia}
\author{Boris N. Chichkov}
\affiliation{Institute of  Quantum Optics, Leibniz Universitat Hannover, Hannover 30167, Germany}
\affiliation{Cluster of Excellence PhoenixD, Leibniz Universitat Hannover, Hannover 30167, Germany}
\affiliation{P.N. Lebedev Physical Institute of the Russian Academy of Sciences, Moscow 119333, Russia}
\author{Andrey B. Evlyukhin}
\email{a.b.evlyukhin@daad-alumni.de}
\affiliation{Institute of  Quantum Optics, Leibniz Universitat Hannover, Hannover 30167, Germany}
\affiliation{Cluster of Excellence PhoenixD, Leibniz Universitat Hannover, Hannover 30167, Germany}
\affiliation{School of Physics and Engineering, ITMO University, St. Petersburg 197101, Russia}
\title{Multipole optimization of light focusing by silicon nanosphere structures}

\begin{abstract}
We investigate the applicability of the coupled multipole model and its modification in the framework of the zero-order Born approximation for modeling of light focusing by finite-size nanostructures of silicon nanospheres, supporting electric and magnetic dipole and quadrupole resonances. The results based on the analytical approximations are verified by comparison with the numerical simulations performed by the T-matrix method. Using the evolutionary algorithm optimization, we apply the developed approach to design the silicon nanosphere metalenses with predefined focusing properties. The obtained results demonstrate a strong optimization potential of the suggested calculation scheme for engineering ultra-thin metalenses.
\end{abstract}
\maketitle

\section{Introduction}
Modern nanophotonics brings a lot of opportunities to control light at the nanoscale. Recent investigations showed that sophisticated designs of nanophotonics components provide beam splitting~\cite{Sell2017},  light trapping~\cite{Lee2017}, objects cloaking~\cite{Deng2016}, multiple and tunable light focusing~\cite{Lin2018}, etc.~\cite{yao2019intelligent} 
However, such remarkable effects are achieved by devices with structural and design complexity.  Optimization of such many-particle systems is a real challenge. To solve this task, different approaches were proposed in the last years \cite{yao2019intelligent}. All of them showed that optimization procedure and its practical realization, including speed and quality, depend strongly on the chosen physical model of the structure and methods of its description. Moreover, the correct choice of the optimization procedure is a decisive condition for a successful result. Therefore, the research activity connected with the development of this field in modern optics is considered to be very important. In this paper, we propose a novel approach to optimization of all-dielectric nanophotonic devices and demonstrate its usability for designing a metalens.

In the proposed method, each structural element is represented by a set of its several multipoles. Multipole decomposition is a method of analyzing the optical properties of subwavelength particles and their arrays in homogeneous or inhomogeneous environments~\cite{miroshnichenko2015substrate, terekhov2017multipolar}. Recent articles  basically focus on stand-alone nanoparticles or infinite periodic arrays. In the last case, the calculation of convergent infinite multipole sums takes into account the interaction between particles~\cite{babicheva_l_appl_phys2021}, but such a method cannot be applied for a finite array. Moreover, the exact numerical simulations can be inefficient for finite structures with many elements due to high waste of computing resources. To simulate electromagnetic interactions in finite particle arrays, several approaches, including multipole methods~\cite{evlyukhin2012collective,babicheva_l_appl_phys2021}, were proposed. Some of them could be realized using the Born approximation (BA), which significantly simplifies the solution process. 

The multipole moments of nanoparticles in an array can be calculated rigorously (i.e., with coupled multipole model~\cite{Babicheva2019}) or approximately by the iterative approach, titled the Born series approach~\cite{Born1926,Hove1954,mishchenko2006multiple,Druger1979,singham1988light, fan2006application, Keller1993,  Bereza2017,Chen2020multilayer,Labani1990}, that applies to the multipole model. In the zero-order BA (zeroth iteration), the multipole moments of nanoparticles are determined only by the incident field, and inter-particle interaction is neglected. In the first-order BA (first iteration), the multipole moments of nanoparticle are determined by the sum of the incident field and the fields of multipole moments, excited in all other particles and calculated in the zero-order BA; in the second-order BA (second iteration), the multipole moments, obtained in the first-order BA, are used, and so on. The Born approximations of different orders has been applied to model the near-field interaction of a microscope tip with a  substrate~\cite{Keller1993,Labani1990} and to calculate the multipole polarizability of a dielectric body~\cite{Druger1979,singham1988light,fan2006application}. Scattering by a pair of dielectric cylinders modelled in the first-order BA demonstrated qualitative agreement with full-wave simulations~\cite{Bereza2017}. In this paper, we follow similar formalism and discuss the application of zero-order Born approximation (ZBA) to the optimization of focusing by metalenses that can greatly simplify and speed up the design of such devices. Our metalenses are composed of silicon (Si) nanospheres supporting electric and magnetic multipole resonances in the visible and near-infrared spectral ranges~\cite{Evlyukhin2010,garcia2011strong,evlyukhin2012demonstration,Kuznetsov2012}.

Similar to a conventional refractive lens, metalens is a light focusing device but consisting of many discrete subwavelength elements~\cite{Khorasaninejad2017,Chen2020flat}. The problem of metalens designing has received a lot of attention in recent years~\cite{lalanne2017metalenses}. This device may substitute refractive lenses in diverse applications where subwavelength sizes and thickness, lightweight, and additional polarization control are essential~\cite{Khorasaninejad2017,Chen2020flat}. In metalenses, the focusing is achieved by adjusting spatial positions or sizes of a metalens structural elements. The prevalent choice of their shape is nanopillar or nanofin~\cite{Chen2020flat}. Recently, the spheres have been considered as metalens structural elements. Using the inverse Mie scattering problem, the positions and sizes of dielectric spheres were optimized, and light focusing in depth-variant discrete helical patterns was demonstrated~\cite{Zhan2019Controlling}. However, in experimental demonstrations, the shape of particles  fabricated by lithography was significantly different from the spherical one. The manufacturing of particles with exact spherical shape by commonly used lithographic methods is practically impossible. On the other hand, there is a fabrication technique called \textit{laser printing of nanoparticles}~\cite{zywietz2014laser} that provides the generation of nanoparticles with an ideal spherical shape, precise positioning, and resonant optical response~\cite{zywietz2015electromagnetic}. Examples of laser printing applications for the fabrication of nanosphere particle arrays with different optical properties, including light focusing, can be found in Ref.~\cite{pique2018laser}.
From a theoretical point of view, spherical nanoparticles are attractive objects for the application of multipole decomposition, because, for them, multipole responses can be considered fully analytically using the Mie theory~\cite{mie1908beitrage,Bohren1983}. 

Following these experimental and theoretical facts, we use spherical nanoparticles as structural elements of metalens. Crystalline silicon is chosen as a particle material due to its high refractive index and low absorption in the visible and near-infrared ranges~\cite{palik1998handbook} that allow the use of metalens in resonant transmission regime. It can also simplify the applicability of zero-order BA because the electromagnetic field in dielectric particles is concentrated predominantly inside them. In general, the proposed multipole approach is not limited by a spherical shape and Si material and can be applied for arbitrary shaped nanoparticles and materials.

One of the key advantages of the multipole model is its direct compatibility with optimization algorithms. Advanced optimization methods showed their high applicability to the designing of photonic nanostructures \cite{yao2019intelligent}. Previously, evolutionary or genetic algorithms were implemented for the development of various nanophotonic devices such as photonic crystals \cite{preble2005two}, waveguide structures \cite{jiang2003parallel}, structures for light focusing~\cite{sanchis2004integrated, huntington2014subwavelength} and localization~\cite{feichtner2012evolutionary}, structural colors~\cite{wiecha2017evolutionary} as well as for solving such a fundamental problem as the inverse scattering problem \cite{huang2007electromagnetic}. In this paper, to implement a light focusing system (metalens) with a desired focal length, we choose an optimization process via Simple Evolutionary Multi-Objective Optimizer (SEMO) algorithm~\cite{Kalyanmoy2002}. This multi-step optimization algorithm adds particles in the structure and engineers its space positions at each step in accordance to one or more optimization goals. The result of the optimization is a metalens design (distribution of particles) with a desired focal length. SEMO-algorithm requires recalculating the multipole moments of each particle a large number of times. Therefore, further simplification of the optimization process is highly desirable. Such simplification can be provided by using the ZBA. In this paper, we formulate the applicability conditions of ZBA and employ ZBA for the optimization process. ZBA simplifies the analysis of physical processes and remarkably reduces the optimization time and consumption of computer resources.

Since the main goal of this work is to develop a new optimization approach, based on the application of ZBA to the multipole light scattering theory, for the design of all-dielectric metalenses using an evolutionary algorithm, the paper is organized as follows. Sec.~\ref{section:CDM_and_ZBA} presents a theoretical multipole model and implementation of the ZBA for the description of the optical response of dielectric nanosphere structures supporting the dipole and quadrupole optical resonances. In Sec.~\ref{section:single_ring}, this theoretical model with the inclusion of the ZBA is applied to investigate the optical properties of ring Si nanosphere structures. Comparison of the obtained results with the results generated by the T-matrix method is used to identify the applicability of the multipole model and the ZBA. Analyzing the strength of the electromagnetic interaction between particles, we find the minimal inter-particle distances when the interaction can be neglected, and the ZBA can be used. The obtained criteria of the ZBA applicability are used in Sec.~\ref{section:optimization} for design and optimization of metalenses by an evolutionary algorithm. Sec.~\ref{section:optimization} demonstrates the optimized c-Si nanosphere metalenses with a desired focal length and high enhancement of the focal intensity. In Sec.~\ref{section:conclusion}, we conclude that, in certain conditions, the optical response of metalenses can be modeled very fast and accurately by the ZBA. This allows fast and efficient optimization of metalenses and other nanophotonic devices.

\section{Coupled multipole model and the  Born approximations}\label{section:CDM_and_ZBA}
All-dielectric metalens is a finite 2D array of dielectric resonant nanoparticles that collectively focus incident light at a point. Refs.~\cite{Wang2017, Khorasaninejad2016, Khorasaninejad2016a, Tanriover2019, Lin2019, PaniaguaDominguez2018} report about the metalenses consisting of differently-shaped nanoparticles for the focusing of a normally incident external field on the metalens axis. In such metalenses, the nanoparticles are arranged in concentric rings. Moreover, the nanoparticles in a single ring are identical because the secondary waves generated by nanoparticles interfere at the focus point constructively if the phase shift of these waves is the same for the nanoparticles equally distant from the metalens axis.

\begin{figure}
\centering
\includegraphics[scale=0.56]{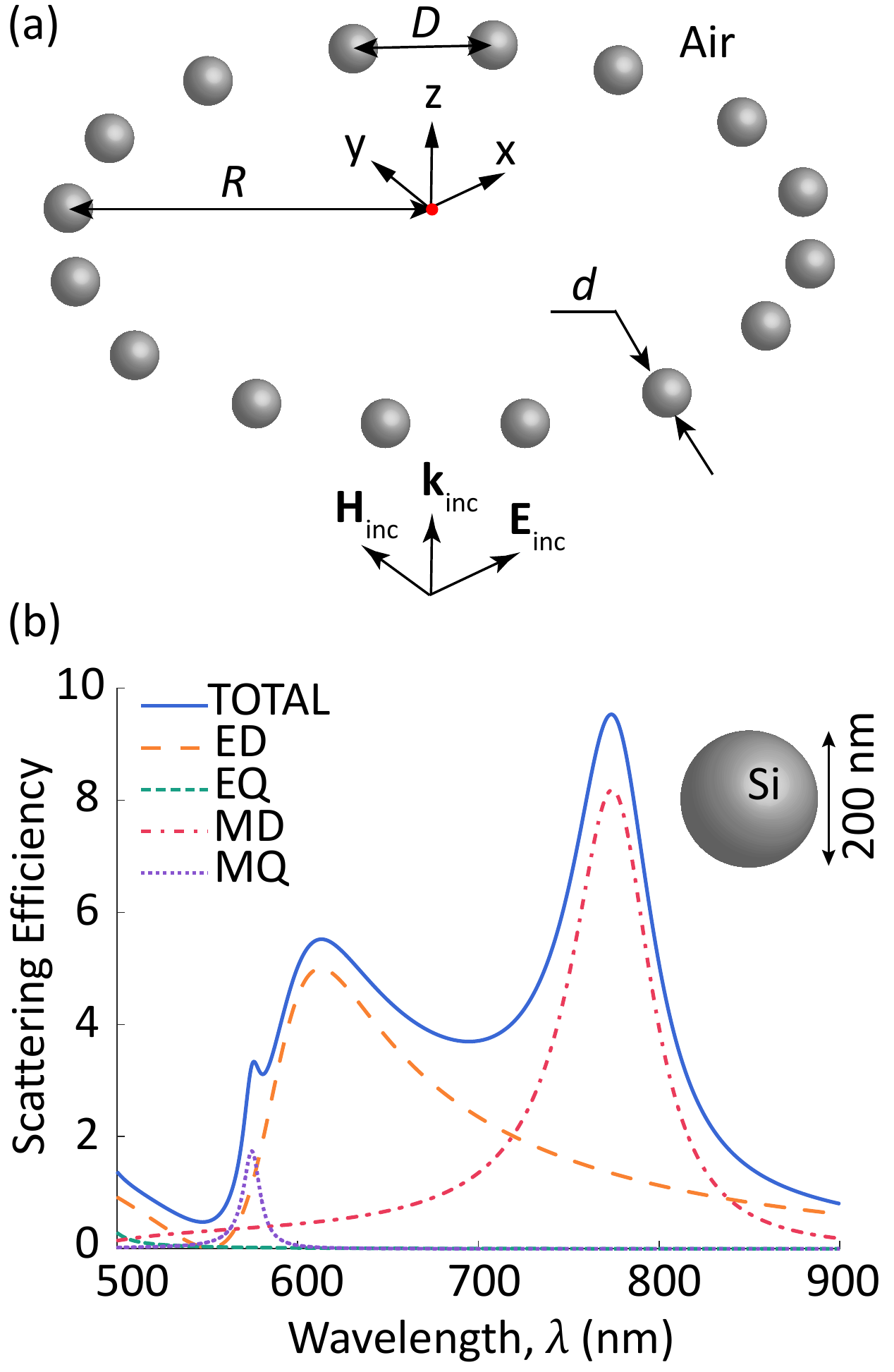}
\caption{(a) Schematic view of the investigated structure which is a ring consisting of Si nanospheres with a diameter $d$ = 200 nm. The ring is placed in $z = 0$ plane. (b) The simulated multipole decomposition of the scattering efficiency for a single Si sphere of diameter $d$ = 200 nm in air. The scattering efficiency and multipole decomposition were calculated using the Mie theory~\cite{Bohren1983}.} 
\label{fig:scheme_and_spectr}
\end{figure}

Due to the above discussion, metalens structures have a coaxial symmetry. Therefore, we start our investigations from a ring of equally separated identical Si nanospheres, shown in Fig.~\ref{fig:scheme_and_spectr}(a). 

To describe the optical response of the ring structure, we consider every nanoparticle as a set of several first electric and magnetic multipoles. The multipoles can be resonantly excited by an external light wave. The type and spectral position of such resonances, known as Mie-resonances~\cite{Bohren1983, jylha2006modeling, zhao2009mie, Evlyukhin2010}, are determined by the particle's shape, size, and material as well as surrounding medium~\cite{evlyukhin2011multipole,Kuznetsov2016, Smirnova2016}. As shown in Fig.~\ref{fig:scheme_and_spectr}(b), the Si nanosphere of the diameter $d=200$ nm supports first dipole and quadrupole resonances in the Si transparency band in the visible and near-IR ranges of the electromagnetic spectrum~\cite{evlyukhin2012demonstration,Kuznetsov2012}. Therefore, its optical properties in this spectral range may be associated with the excitation of only the first four major multipoles: magnetic dipole (MD), electric dipole (ED), magnetic quadrupole (MQ), and electric quadrupole (EQ). Note that even in case when the dipole approximation is sufficient for a single nanoparticle, the inter-particle coupling in nanoparticle structures can lead to the excitation of their higher-order multipoles, which requires their consideration in the developed model~\cite{babicheva2018metasurfaces,Babicheva2019}. Interference between the waves irradiated by these multipoles leads to different optical effects~\cite{liu2017multipolar}. 

Using the results in Fig. \ref{fig:scheme_and_spectr}(b), we can represent each nanosphere in a ring structure as a set of two dipoles and two quadrupoles located at one point coinciding with the particle center. An analytical approach for the investigation of optical properties of such systems has been recently developed \cite{Babicheva2019}. In this paper, we name it the Coupled Multipole Model (CMM). The vectors of electric $\mathbf{p}^j$ and magnetic $\mathbf{m}^j$ dipoles and tensors of electric $\hat{Q}^j$ and magnetic $\hat{M}^j$ quadrupoles of the \textit{j}-th nanosphere located at $\mathbf{r}_j$ ($j = 1, 2, 3, ... ,N$, where $N$ is the total number of nanospheres) are determined by the local electric $\mathbf{E}_{\mathrm{loc}}(\mathbf{r}_j)$ and magnetic $\mathbf{H}_{\mathrm{loc}}(\mathbf{r}_j)$ fields acting on the nanosphere, respectively \cite{Babicheva2019}:
\begin{gather}
\label{el_dip_loc}
\mathbf{p}^{j} = \alpha_{p} \mathbf{E}_{\mathrm{loc}}\left(\mathbf{r}_j\right), \quad
\mathbf{m}^{j} = \alpha_{m} \mathbf{H}_{\mathrm{loc}}\left(\mathbf{r}_j\right), \\
\label{el_quad_loc}
\hat{Q}^{j} = \frac{\alpha_{Q}}{2}\left[\bm{\nabla}_j \otimes \mathbf{E}_{\mathrm{loc}}\left(\mathbf{r}_j\right) + \left(\bm{\nabla}_j \otimes \mathbf{E}_{\mathrm{loc}}\left(\mathbf{r}_j\right)\right)^{T}\right], \\
\label{mag_quad_loc}
\hat{M}^{j} = \frac{\alpha_{M}}{2}\left[\bm{\nabla}_j \otimes \mathbf{H}_{\mathrm{loc}}\left(\mathbf{r}_j\right) + \left(\bm{\nabla}_j \otimes \mathbf{H}_{\mathrm{loc}}\left(\mathbf{r}_j\right)\right)^{T}\right],
\end{gather}
where $^T$ denotes the transpose operation; $\otimes$ denotes the dyadic product; $\bm{\nabla}_j$ is the nabla operator with respect to $\mathbf{r}_j$; $\alpha_{p}$, $\alpha_{m}$, $\alpha_{Q}$, and $\alpha_{M}$ are the ED, MD, EQ, and MQ polarizabilities of a dielectric sphere, respectively. We consider homogeneous isotropic dielectric nanospheres, when anisotropic and bianisotropic properties \cite{asadchy2018bianisotropic,evlyukhin2020bianisotropy} do not appear. 

For the \textit{j}-th nanosphere, the local electric [magnetic] field is a superposition of the incident electric $\mathbf{E}_{\mathrm{inc}}(\mathbf{r}_j)$ [magnetic  $\mathbf{H}_{\mathrm{inc}}(\mathbf{r}_j)$] field, the electric $\mathbf{E}_{p}'\left(\mathbf{r}_{j}\right)$ [magnetic $\mathbf{H}_{p}'\left(\mathbf{r}_{j}\right)$] field generated by all EDs of the array except $\mathbf{p}^j$, the field $\mathbf{E}_{m}'\left(\mathbf{r}_{j}\right)$ [$\mathbf{H}_{m}'\left(\mathbf{r}_{j}\right)$] of all MDs except $\mathbf{m}^j$, the field $\mathbf{E}_{Q}'\left(\mathbf{r}_{j}\right)$ [$\mathbf{H}_{Q}'\left(\mathbf{r}_{j}\right)$] of all EQs except $\hat{Q}^j$ and the field $\mathbf{E}_{M}'\left(\mathbf{r}_{j}\right)$ [$\mathbf{H}_{M}'\left(\mathbf{r}_{j}\right)$] of all MQs except $\hat{M}^j$:
\begin{eqnarray}
\mathbf{E}_{\mathrm{loc}}\left(\mathbf{r}_j\right)&=& \mathbf{E}_{\mathrm{inc}}\left(\mathbf{r}_j\right) + {\mathbf{E}}_{p}'\left(\mathbf{r}_{j}\right)+{\mathbf{E}}_{m}'\left(\mathbf{r}_{j}\right) \nonumber\\
&& + {\mathbf{E}}_{Q}'\left(\mathbf{r}_{j}\right)+{\mathbf{E}}_{M}'\left(\mathbf{r}_{j}\right), 
\label{el_field_loc}\\
\mathbf{H}_{\mathrm{loc}}\left(\mathbf{r}_j\right)& =& \mathbf{H}_{\mathrm{inc}}\left(\mathbf{r}_j\right) + {\mathbf{H}}_{p}'\left(\mathbf{r}_{j}\right)+{\mathbf{H}}_{m}'\left(\mathbf{r}_{j}\right)\nonumber\\ 
&&+ {\mathbf{H}}_{Q}'\left(\mathbf{r}_{j}\right)+{\mathbf{H}}_{M}'\left(\mathbf{r}_{j}\right).
\label{mag_field_loc}
\end{eqnarray}

The expressions for electric ${\mathbf{E}'}$ and magnetic ${\mathbf{H}'}$ fields generated by dipole (quadrupole) moments are determined using the free space dipole (quadrupole) Green's tensors:
\begin{align}
\label{eq:multipole_fields}
\begin{aligned}
{\mathbf{E}}_{p}'\left(\mathbf{r}_{j}\right)& = \frac{k_{0}^{2}}{\varepsilon_{0}} \sum\limits_{l = 1, l \neq j}^{N} \hat{G}_{j l}^{p} \mathbf{p}^l, \\
{\mathbf{H}}_{p}'\left(\mathbf{r}_{j}\right) &= \frac{c k_{0}}{\mathrm{i}} \sum\limits_{l = 1, l \neq j}^{N}\mathbf{g}_{jl} \times \mathbf{p}^{l},\\
{\mathbf{E}}_{m}'\left(\mathbf{r}_{j}\right) &=
\frac{\mathrm{i} k_{0}}{c \varepsilon_{0}} \sum\limits_{l = 1, l \neq j}^{N} \mathbf{g}_{jl} \times \mathbf{m}^{l}, \\
{\mathbf{H}}_{m}'\left(\mathbf{r}_{j}\right) &= k_{S}^{2} \sum\limits_{l = 1, l \neq j}^{N} \hat{G}_{j l}^{p} \mathbf{m}^{l}, \\
{\mathbf{E}}_{Q}'\left(\mathbf{r}_{j}\right) &= \frac{k_{0}^{2}}{\varepsilon_{0}} \sum\limits_{l = 1, l \neq j}^{N} \hat{G}_{j l}^{Q}\left(\hat{Q}^{l} \mathbf{n}_{l j}\right), \\
{\mathbf{H}}_{Q}'\left(\mathbf{r}_{j}\right) &= \frac{c k_{0}}{\mathrm{i}} \sum\limits_{l = 1, l \neq j}^{N} \mathbf{q}_{jl} \times \left(\hat{Q}^{l} \mathbf{n}_{l j}\right), \\
{\mathbf{E}}_{M}'\left(\mathbf{r}_{j}\right) &= 3\frac{\mathrm{i} k_{0}}{c \varepsilon_{0}} \sum\limits_{l = 1, l \neq j}^{N} \mathbf{q}_{jl} \times \left(\hat{M}^{l} \mathbf{n}_{l j}\right),\\
{\mathbf{H}}_{M}'\left(\mathbf{r}_{j}\right)&= 3 k_{S}^{2} \sum\limits_{l = 1, l \neq j}^{N} \hat{G}_{j l}^{Q} \left(\hat{M}^{l} \mathbf{n}_{l j}\right),
\end{aligned}
\end{align}
where i is the imaginary unit; $\varepsilon_0$ is the vacuum dielectric constant; $c$ is the vacuum speed of light; $k_0$ is the wavenumber of incident wave in vacuum,  and $k_{S} = k_0 \sqrt{\varepsilon_{S}}$; $\varepsilon_{S}$ is the relative dielectric permittivity of the host medium (in this paper, we consider $\varepsilon_{S}$ = 1); $\mathbf{n}_{lj} = (\mathbf{r}_j - \mathbf{r}_l)/|\mathbf{r}_j - \mathbf{r}_l|$ is the unit vector directed from $\mathbf{r}_l$ to $\mathbf{r}_j$ (here $\mathbf{r}_j$ is the field calculation point, and $\mathbf{r}_l$ is the position of the field source); $\hat{G}_{j l}^{p} \equiv \hat{G}^{p}(\mathbf{r}_j, \mathbf{r}_l)$ and $\hat{G}_{j l}^{Q} \equiv \hat{G}^{Q}(\mathbf{r}_j, \mathbf{r}_l)$ are the dyadic Green's functions of a point electric dipole and quadrupole in free space, respectively. The analytical expressions for Green's functions $\hat{G}_{j l}^{p}$, and $\hat{G}_{j l}^{Q}$ can be found in Ref.~\cite{Babicheva2019} [see Eqs.(10), (11)]. Auxiliary vectors $\mathbf{g}_{jl}$ and $\mathbf{q}_{jl}$ connect with dyadic Green's functions as following:
\begin{align}
\label{g_definition}
 \mathbf{g}_{jl} \times \mathbf{p}^j &= \bm{\nabla}_j \times \hat{G}_{jl}^{p} \mathbf{p}^j, \\ 
 \label{q_definition}
\mathbf{q}_{jl} \times \left( \hat{Q}^{l} \mathbf{n}_{l j}\right) &= \bm{\nabla}_j \times \hat{G}_{jl}^{Q} \left(\hat{Q}^{l}\mathbf{n}_{l j}\right).
\end{align}
For analytical expressions of vectors $\mathbf{g}_{jl}$ and $\mathbf{q}_{jl}$, see Eqs. (17), (18) in Ref.~\cite{Babicheva2019}.

Total electric and magnetic field at any observation point $\mathbf{r}$ are determined as a superposition of the incident external field and the fields generated by all dipoles and quadrupoles in the system:
\begin{eqnarray}
\label{total_el_field}
\mathbf{E}\left(\mathbf{r}\right)& =& \mathbf{E}_{\mathrm{inc}}\left(\mathbf{r}\right) + \frac{k_0^2}{\varepsilon_0}\sum\limits_{j=1}^{N} \left\{\hat{G}^{p}(\mathbf{r},\mathbf{r}_j) \mathbf{p}^{j}\right. \nonumber\\
&&+ \frac{\mathrm{i}}{c k_{0}}\left[\mathbf{g}(\mathbf{r},\mathbf{r}_j) \times \mathbf{m}^{j} \right] +\hat{G}^{Q}(\mathbf{r},\mathbf{r}_j)\left(\hat{Q}^{j} \mathbf{n}_j\right)\nonumber\\
&&\left.+ \frac{3 \mathrm{i}}{c k_{0}}\left[\mathbf{q}(\mathbf{r},\mathbf{r}_j) \times\left(\hat{M}^{j}  \mathbf{n}_{j}\right) \right]\right\}, \\
\label{total_mag_field}
\mathbf{H}\left(\mathbf{r}\right)& =& \mathbf{H}_{\mathrm{inc}}\left(\mathbf{r}\right) + k_0^2 \sum\limits_{j=1}^{N}\left\{ \frac{c}{\mathrm{i} k_{0}}\left[\mathbf{g}(\mathbf{r},\mathbf{r}_j) \times \mathbf{p}^{j}\right]\right.\nonumber\\
&& + \varepsilon_{S} \hat{G}^{p}(\mathbf{r},\mathbf{r}_j) \mathbf{m}^{j} +  \frac{c}{\mathrm{i} k_{0}}\left[\mathbf{q}(\mathbf{r},\mathbf{r}_j) \times \left(\hat{Q}^{j} \mathbf{n}_{j}\right) \right]\nonumber\\
&&\left.+ 3 \varepsilon_{S} \hat{G}^{Q}(\mathbf{r},\mathbf{r}_j)\left(\hat{M}^{j} \mathbf{n}_{j}\right) \right\},
\end{eqnarray}
where $\mathbf{n}_{j} = (\mathbf{r} - \mathbf{r}_j)/|\mathbf{r} - \mathbf{r}_j|$.
Multipole polarizabilities of a single sphere are expressed in terms of the scattering coefficients $a_1$, $b_1$, $a_2$, and $b_2$ from the Mie-theory~\cite{Bohren1983} in following manner~\cite{Babicheva2019}:
\begin{align}
\label{polarizabilities}
\begin{aligned}
    &\alpha_{p}=\mathrm{i} \frac{6 \pi \varepsilon_{0} \varepsilon_{S}}{k_{S}^{3}} a_{1}, \quad &
    &\alpha_{m}=\mathrm{i} \frac{6 \pi}{k_{S}^{3}} b_{1}, \\
    &\alpha_{Q}=\mathrm{i} \frac{120 \pi \varepsilon_{0} \varepsilon_{S}}{k_{S}^{5}} a_{2}, \quad &
    &\alpha_{M}=\mathrm{i} \frac{40 \pi}{k_{S}^{5}} b_{2}.
\end{aligned}
\end{align}

After substitution of local fields (\ref{el_field_loc})-(\ref{mag_field_loc}) into multipole moments (\ref{el_dip_loc})-(\ref{mag_quad_loc}), we obtain the linear system of equations for the calculation of dipole and quadrupole moments of all nanospheres in the structure. The detailed system is explicitly written in Ref. \cite{Babicheva2019}, Eq. (25). Here we write this system in a matrix form more suitable for numerical calculations:
\begin{gather}
\label{impl_system}
    \mathbf{Y} = \mathbf{Y}_0 + \hat{\mathbf{V}}\mathbf{Y},
\end{gather}
where $\mathbf{Y}$ and $\mathbf{Y}_0$ are the 24$N$ dimensional supervectors composed of dipole and quadrupole moments:
\begin{eqnarray}
\label{Y_vector}
\mathbf{Y}& =&\left[p^1_x, \hdots, p_z^N, m_x^1, \hdots, m_z^N, Q^1_{xx},  \hdots, Q^N_{zz},\right.\nonumber\\
&&\left. M^1_{xx}, \hdots, M^N_{zz} \right]^{T}, \\
\label{Y_0_vector}
\mathbf{Y}_0 &=&\left[p^1_{0,x}, \hdots, p^N_{0,z}, m^1_{0,x}, \hdots, m_{0,z}^N, Q^1_{0,xx}, \hdots, Q^N_{0,zz},\right.\nonumber\\
&& \left. M^1_{0,xx}, \hdots, M^N_{0,zz} \right]^{T}.
\end{eqnarray}
$\mathbf{Y}$ is the supervector of the coupled multipole moments, taking into account the interaction of particles; $\mathbf{Y}_0$ is the supervector of multipole moments excited only by the incident wave. These moments (indicated by subscript 0) are expressed as (\ref{el_dip_loc})-(\ref{mag_quad_loc}) by replacing local fields with the incident fields. $\hat{\mathbf{V}}$ is the block-matrix of interaction between multipoles of dimension $24N\times24N$, in general. Note that symmetrical and traceless properties of the quadrupole tensors \cite{evlyukhin2012collective} can reduce the number of unknown variables and the dimension of matrix $\hat{\mathbf{V}}$ in Eq. (\ref{impl_system}). Explicit presentation of matrix $\hat{\mathbf{V}}$ is followed from Eqs. (\ref{el_dip_loc})-(\ref{mag_quad_loc}) written for multipoles of all particles in the structure. The dipole moments of nanoparticle are determined by local fields (\ref{el_dip_loc}) while the quadrupole moments  are determined by derivatives of local fields (\ref{el_quad_loc}), (\ref{mag_quad_loc}). The local fields determined by Eqs. (\ref{el_field_loc})-(\ref{eq:multipole_fields}) involve the fields of multipoles expressed through Green's functions, and the vectors $\mathbf{g}$ and $\mathbf{q}$ (see Eqs. (10), (11), (17), (18) in~\cite{Babicheva2019}). Expressions for their derivatives determining the quadruple moments are presented in Appendix~\ref{section:green} by Eqs. (\ref{eq:d_Gpp})-(\ref{eq:d_GQM}). 

Formally, the CMM-solution of Eq. (\ref{impl_system}) can be express as following:
\begin{gather}
\label{exact_sol}
    \mathbf{Y} = (\hat{\mathbf{I}} - \hat{\mathbf{V}})^{-1}\mathbf{Y}_0,
\end{gather}
where $\hat{\mathbf{I}}$ is the corresponding identity matrix. However, applying the Born series method, the system~(\ref{impl_system}) can be solved by an iteration procedure, which saves the computation time for large amounts of nanoparticles. The Born series expansion of (\ref{exact_sol}) can be obtained by expanding the matrix $(\hat{\mathbf{I}} -\hat{\mathbf{V}})^{-1}$ in powers of the matrix $\hat{\mathbf{V}}$:
\begin{gather}
\label{born_series}
    \mathbf{Y} = \mathbf{Y}_0 + \hat{\mathbf{V}}\mathbf{Y}_0 + \hat{\mathbf{V}}^2\mathbf{Y}_0 + \hdots.
\end{gather}
Expansion (\ref{born_series}) allows us to write successive Born approximations to the solution (\ref{exact_sol}). Neglecting interactions between the particles, a system solution can be written in the zero-order Born approximation (ZBA): 
\begin{gather}
\label{ZBA}
    \mathbf{Y} = \mathbf{Y}_0.
\end{gather}
The $n$-th order Born approximation is expressed through the previous-order one by the relationship:
\begin{gather}
\label{born_approximation}
    \mathbf{Y}_n = \mathbf{Y}_0 + \hat{\mathbf{V}}\mathbf{Y}_{n-1}.
\end{gather}
Thus, the Born series approach, including ZBA, can be applied when the inter-particle interaction is weak. On the contrary, the Born series (\ref{born_series}) diverge and can not provide correct results at the condition of configuration resonances \cite{Keller1993} when $\det(\hat{\mathbf{I}} - \hat{\mathbf{V}}) = 0$. In this case, the electromagnetic coupling between nanoparticles in the structures is very strong. Note that in this paper we consider only systems that do not support such configuration resonances.

The advantages of the ZBA are following: analytical simplicity, and saving of computing resources and calculation time for many-particle structures. For the structure of $N$ particles, the CMM-solution of the system (\ref{exact_sol}) needs $O(N^3)$ of computational time units, the $n$-th order Born approximation needs $O(nN^2)$, while the ZBA-solution needs only $O(N)$~\cite{Cormen2009}.

\section{Single ring modeling and criteria of the CMM and ZBA applicability}\label{section:single_ring}
Using the method described in Sec. \ref{section:single_ring}, we numerically simulate the optical response of the ring composed of $N$ Si nanospheres with the diameter $d=200$ nm. The ring is placed in the $xy$-plane ($z=0$), and the origin of the Cartesian coordinate system is located at the ring center [Fig. \ref{fig:scheme_and_spectr}(a)]. The coordinates of the $j$-th sphere in the ring are $[R\cos{(2\pi (j-1)/N)}, R\sin{(2\pi (j-1)/N)}, 0]$, where $R$ is the ring radius, and $j = 1, 2, \hdots, N$. Thus, the distance to the nearest neighbors is the same for every particle in the structure. According to Fig.~\ref{fig:scheme_and_spectr}(a), the ring is illuminated by a normally incident plane wave traveling along the $z$-axis ($\mathbf{k}_{\mathrm{inc}} = k_S \hat{\mathbf{z}}$) with the following  electric $\mathbf{E}_{\mathrm{inc}}(\mathbf{r}) = E_0 e^{\mathrm{i} k_S z} \hat{\mathbf{x}}$ and magnetic $\mathbf{H}_{\mathrm{inc}}(\mathbf{r}) = (E_0/Z) e^{\mathrm{i} k_S z}  \hat{\mathbf{y}} $ fields. Here, $\hat{\mathbf{x}}$, $\hat{\mathbf{y}}$, and $\hat{\mathbf{z}}$ are the unit vectors along the $x$-, $y$-, and $z$-axis, respectively; $Z = \sqrt{\mu_0/\varepsilon_0\varepsilon_S}$ is the medium impedance. As the surrounding medium is air ($\varepsilon_S = 1$), then $k_S = k_0 = 2\pi/\lambda$, where $\lambda$ is the free-space wavelength. All simulations take into account dispersion of Si refractive index and extinction coefficient \cite{aspnes1983dielectric}.

\begin{figure}
    \centering
    \includegraphics[scale=0.7]{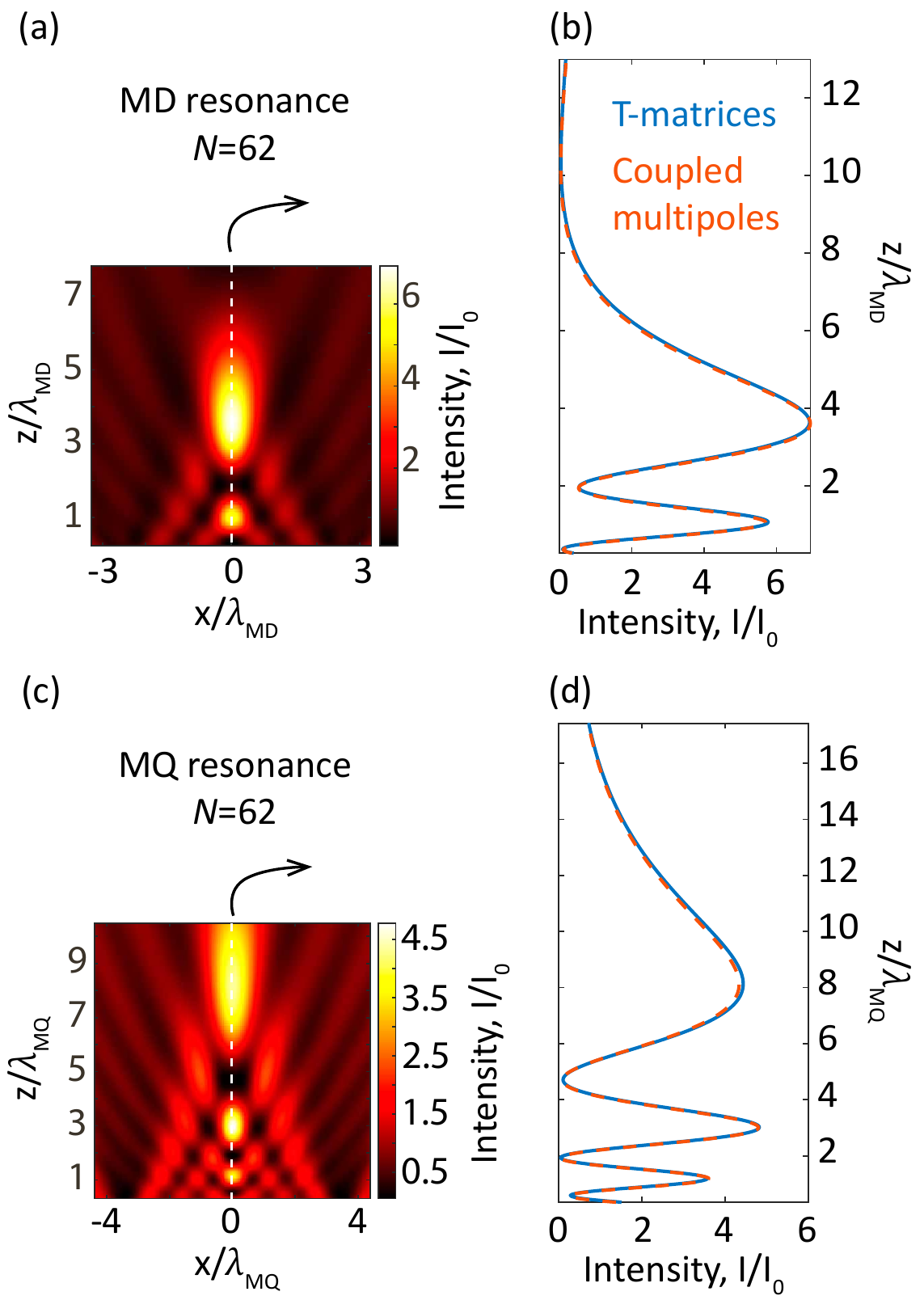}
    \caption{Normalized  intensity  calculated using the CMM  for a single ring [see Fig. \ref{fig:scheme_and_spectr}(a)] of radius $R$ = 2 $\mu$m and number of particles $N = 62$ at the wavelength (a) $\lambda_{\mathrm{MD}} = 770$ nm of the MD resonance, and (c) $\lambda_{\mathrm{MQ}} = 574$ nm of the MQ resonance. The intensities  (a) and (c) illustrate that the ring of 62 equally separated nanospheres focuses light at both resonant wavelengths. (b) and (d) The simulated normalized intensity along the $z$-axis from the T-matrix method~\cite{Egel2017celes} (blue solid lines) and CMM~(\ref{exact_sol}) (red dashed lines). The normalization factor is the intensity of the incident plane wave.}
    \label{fig:multipole_model}
\end{figure}

To verify the coupled multipole model (CMM), we compare the electromagnetic field intensity calculated using the multi-sphere T-matrix code by A. Egel \textit{et al}.~\cite{Egel2017celes} and the CMM-solution (\ref{exact_sol}) of the system (\ref{impl_system}) written for our multipole model. In the latter approach, the multipole moments are calculated using (\ref{exact_sol}) with subsequent substitution in the expressions for total electric and magnetic fields (\ref{total_el_field}). The CMM-solution of Eq.~(\ref{impl_system}) and calculations of fields were realized in the MATLAB~\cite{MATLAB_2019}. Using the total electric and magnetic fields, we calculate the normalized intensity of the electromagnetic field by the following formula: $\mathrm{I}(\mathbf{r})/\mathrm{I}_0 = (|\mathbf{E}(\mathbf{r})|^2 + Z^2 |\mathbf{H}(\mathbf{r})|^2)/2|E_0|^2$, where $Z$ is  the vacuum wave impedance. Figure~\ref{fig:multipole_model} shows the comparison of results for the normalized intensity profiles at wavelengths of MQ resonance (574 nm) and MD resonance (770 nm) near the ring of radius $R$ = 2 $\mu$m and number of particles $N = 62$.

The applicability of the dipole approximation to systems of metallic (plasmonic) nanospheres has been discussed, for example, in Refs.~\cite{romero2006plasmons, Zou2004}. It has been shown that a significant divergence between the full-wave simulation and the coupled dipole model is observed when the distance between the centers of the particles $D$ is less than or equal to their doubled diameter, i.e., $D \leq 2d$. 

Let us find the criterion for the ring structure of dielectric spheres in the coupled dipole-quadrupole model. In Fig.~\ref{fig:multipole_model} the number of spheres $N = 62$ corresponds to $D \approx d$ (the neighboring particles almost touch each other) when the interaction between particles should be strongest. For magnetic dipole and quadrupole resonances, the intensity profiles calculated using the CMM and T-matrix method agree well, as shown in Figs.~\ref{fig:multipole_model}(b) and \ref{fig:multipole_model}(d).
In the case of metalens designing, we should investigate the calculation accuracy of the structure's focal length and focal intensity that are fundamental parameters of any metalens, determining its functional properties. Let us define as a focal length of the ring the distance between the global intensity maximum on the $z$-axis and the ring plane. The relative error of the ring's focal length in the CMM (comparing the values obtained by the CMM and T-matrix methods): 0.71\% for MD and 1.15\% for MQ resonances; the relative error of the CMM focal intensity value: 0.33\% for MD and 0.39\% for MQ resonances. Thus, the model of coupled multipoles (dipoles and quadrupoles) is applicable for almost any distance between dielectric spherical particles (i.e., for $D > d$) at dipole and quadrupole resonances of a single particle.

\begin{figure}
    \centering
   \includegraphics[scale=0.55]{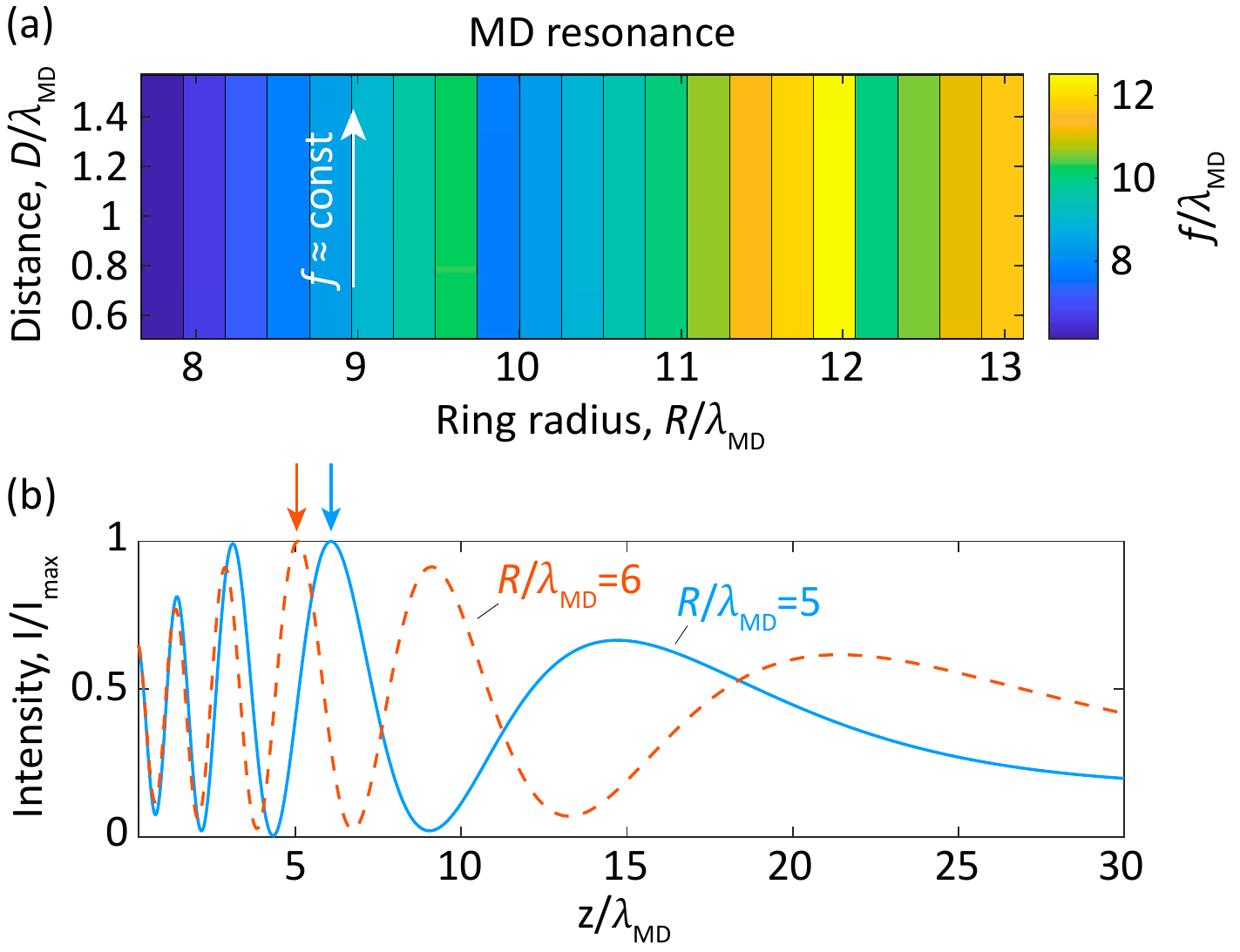}
    \caption{(a) Focal length calculated using the ZBA as a function of the ring radius and inter-particle (center-center) distance at wavelength of the single-partice MD resonance. The focal length is defined as a distance between the ring plane and the point on the $z$-axis corresponding to the global maximum of light intensity. (b) Intensity (normalized on the maximum value) of the fields generated by two rings ($N = 32$) along the $z$-axis calculated at the MD resonance in the ZBA [Eq.~(\ref{intensity_z_axis}) in Appendix~\ref{section:intensity}]. Red and blue arrows in (b) indicate the focal lengths of rings with the radius $R/\lambda_{\rm MD} = 6$ and $R/\lambda_{\rm MD} = 5$, respectively. All dimensional values in (a) and (b) are normalized by the MD resonance wavelength $\lambda_{\rm MD} = 770$ nm.}
    \label{Focus_length}
\end{figure}

Let us now consider application of the ZBA. Figures \ref{fig:multipole_model}(a) and \ref{fig:multipole_model}(c) show that the rings exhibit focusing properties, creating hot spots. The intensity profile of the nanosphere ring is similar to the picture of the Fresnel diffraction on a circle hole structure (see, e.g., Fig. 4 in Ref. \cite{Gillen2004diffraction}), but, in contrast to a circle-shaped hole, every nanoparticle in the ring structures supports a resonant response. Figure~\ref{Focus_length}(a) demonstrates the focal length (calculated in the ZBA) of the ring with varying radius $R$ and inter-particle distance $D$. One can see that at the MD resonance, the focal length does not change for the ring with a fixed radius and a different number of particles (i.e., different inter-particle distance). Thus, the focal length, in particular, and spatial positions of intensity peaks on the $z$-axis are practically independent of the number of particles in the ring, but these quantities depend on the ring radius [see Fig.~\ref{Focus_length}(a)] and wavelength [see Fig.~\ref{fig:multipole_model}]. Note that the number of nanoparticles determines the focal intensity enhancement (see Eq.~(\ref{intensity_z_axis}) in Appendix~\ref{section:intensity} for expression of the intensity in the ZBA).

Figure~\ref{Focus_length}(a) also shows that the focal length is non-monotonically changed with the increasing ring radius. 
Such behavior is explained by the following discussion. The constructive interference between the external incident wave and the scattered waves yields the local intensity peaks along the $z$-axis [see Eq.~(\ref{intensity_z_axis}) in Appendix~\ref{section:intensity}]. These peaks move away from the ring plane as the ring's radius increases, as shown in Fig.~\ref{Focus_length}(b). As a result, with a larger radius, the global intensity maximum (i.e., the focus) may be closer to the ring. When this happens, the focal length decreases as the radius of the ring increases [see focal lengths of the rings in Fig.~\ref{Focus_length}(b)]. Similar behavior (not shown in Figures)  of the focal length is observed at the MQ resonance.

\begin{figure}
    \centering
    \includegraphics[scale=0.415]{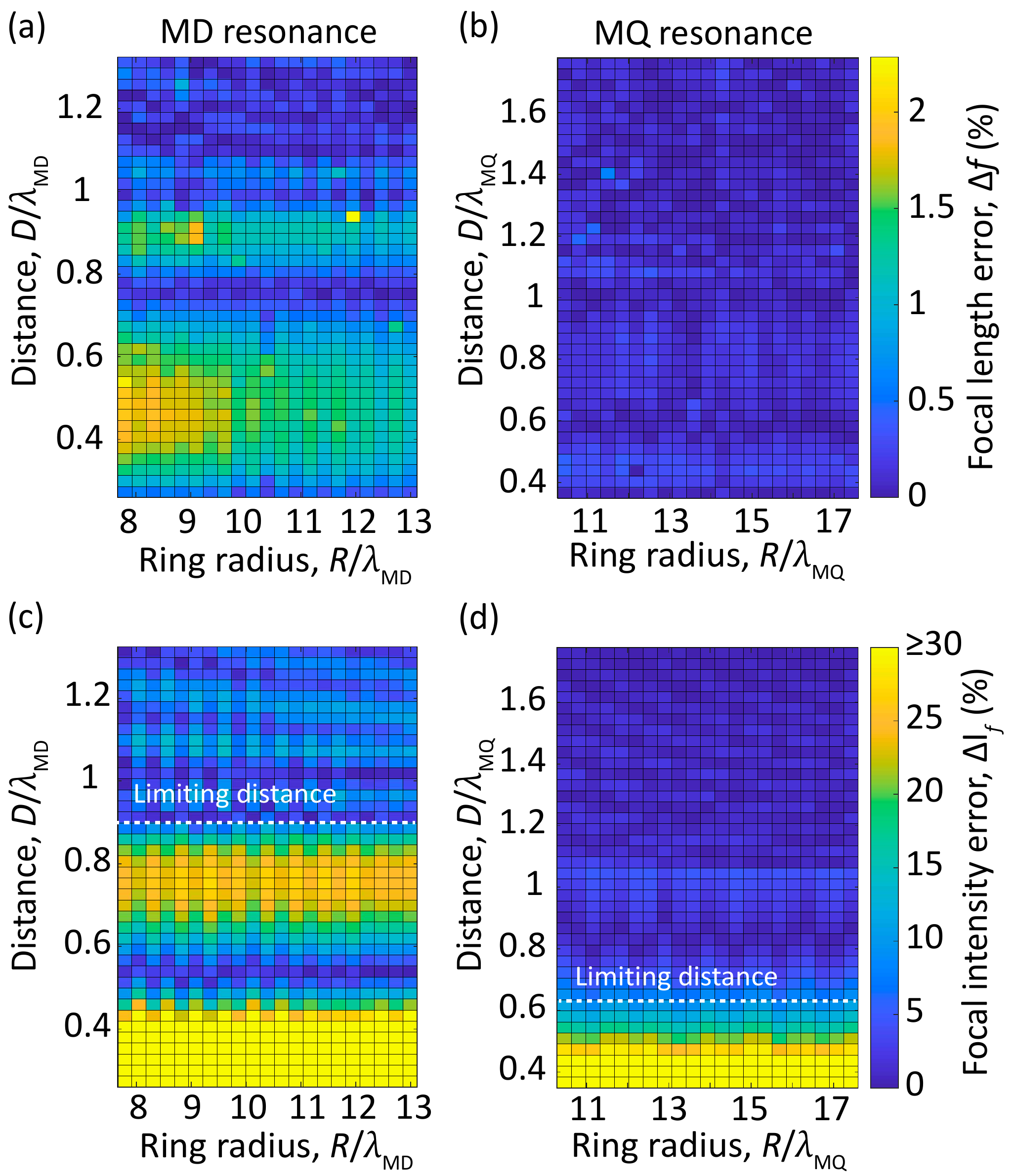}
    \caption{Focal length error  $\Delta f$ of the ZBA as a function of the ring radius and inter-particle (center-center) distance at wavelengths of (a) the MD ($\lambda_{\mathrm{MD}} = 770$ nm) and (b) the MQ ($\lambda_{\mathrm{MQ}} = 574$ nm) resonances.  Focal intensity error $\Delta \mathrm{I}_f$ of ZBA  for (c) MD and (d) MQ resonances. The white dashed lines indicate the {\it limiting distances}: for all larger distances, the corresponding errors of ZBA are the order or less than  10\%.}
    \label{fig:MaxInt_and_Focus_Errors_vs_D_and_RR}
\end{figure}

As we mentioned above, the optimization process will be performed in the framework of ZBA. Therefore, we should clarify the accuracy ofthe ZBA for a single ring. Let us now estimate relative deviations (errors) $\Delta V$ between the quantities obtained in the framework of CMM ($V^{\rm (CMM)}$) and ZBA ($V^{\rm (ZBA)}$): 
\begin{equation}
    \Delta V =\frac{|V^{\rm (CMM)}-V^{\rm (ZBA)}|}{V^{\rm (CMM)}}\times100 \%,
\end{equation}
where $V^{\rm (CMM)}$ and $V^{\rm (ZBA)}$ are certain values of the quantity $V$ calculated using CMM [see Eq. (\ref{exact_sol})] and ZBA [see Eq. (\ref{ZBA})] approaches, respectively. Figure~\ref{fig:MaxInt_and_Focus_Errors_vs_D_and_RR} shows the errors of the ring focal length $\Delta f$ and the ring focal intensity $\Delta \mathrm{I}_f$ calculated as functions of the inter-particle distance and ring radius. 
The ZBA accurately determines the focal length for all considered ranges of parameters: the error is less than 2.5\% for MD resonance and 1\% for MQ resonance [see Figs.~\ref{fig:MaxInt_and_Focus_Errors_vs_D_and_RR}(a) and \ref{fig:MaxInt_and_Focus_Errors_vs_D_and_RR}(b)]. It means that, at resonance frequencies, ZBA is applicable for calculating the focal length of ring structures for any inter-particle distance $D > d$. The error for the intensity at the focal point is presented in Figs.~\ref{fig:MaxInt_and_Focus_Errors_vs_D_and_RR}(c) and \ref{fig:MaxInt_and_Focus_Errors_vs_D_and_RR}(d). In contrast to the focal length definition, the accuracy of intensity calculations in the ZBA is lower in the considered ranges of parameters [see Figs.~\ref{fig:MaxInt_and_Focus_Errors_vs_D_and_RR}(c) and \ref{fig:MaxInt_and_Focus_Errors_vs_D_and_RR}(d)]. 

Substantially, beginning from a certain inter-particle distance, the error of the focal intensity is quite low. Hence, we can determine a lower limit on the distance between particles starting from which the error of focal intensity in the ZBA $\Delta \mathrm{I}_f \lesssim 10$\% for all ring radii. In Figs.~\ref{fig:MaxInt_and_Focus_Errors_vs_D_and_RR}(c) and \ref{fig:MaxInt_and_Focus_Errors_vs_D_and_RR}(d), this distance is indicated by the white dashed line and named as the limiting distance. The limiting distance is equal to 0.92$\lambda_{\mathrm{MD}}$ for MD resonance (770 nm), and 0.66$\lambda_{\mathrm{MQ}}$ for MQ resonance (574 nm). In terms of sphere diameter ($d$ = 200 nm), the limiting distance is equal to $3.54d$ and $1.9d$, respectively. Further, we take into account these quantities in the optimization procedure. Note that the differences between the error distributions related to MD and MQ resonances in Fig.~\ref{fig:MaxInt_and_Focus_Errors_vs_D_and_RR} have resulted from differences in radiation (scattering) directivity of the MD and MQ multipole sources. 

\section{Metalens design and optimization using  the  ZBA}\label{section:optimization}
Using the results of the previous section,  apply the ZBA for realization of an evolutionary algorithm for design and optimization of metalenses with predefined properties. A polarization-independent metalens should be composed of various concentric rings of nanoparticles that resonantly scatter incident waves. In this configuration, we know that each ring produces a set of intensity peaks on the optical axis ($z$-axis) due to interference between the incident and scattered waves (in Appendix~\ref{section:intensity}, see analytical expression (\ref{intensity_z_axis}) of the $z$-axis intensity in the framework of the ZBA). The number of these peaks increases and their positions shift along the optical axis with the increasing ring radius. Consequently, in a structure consisting of several rings, the strongest focusing effect can be reached if all rings have intensity peak at the same point. But this is not a sufficient condition; it is also important that the phases of fields generated by all rings at the position of these peaks should be the same. The optimization procedure defines the optimum number of rings and their parameters, such as the ring radius and the number of particles, to get a focus in a certain point.

\begin{figure}
\centering
\includegraphics[scale=0.36]{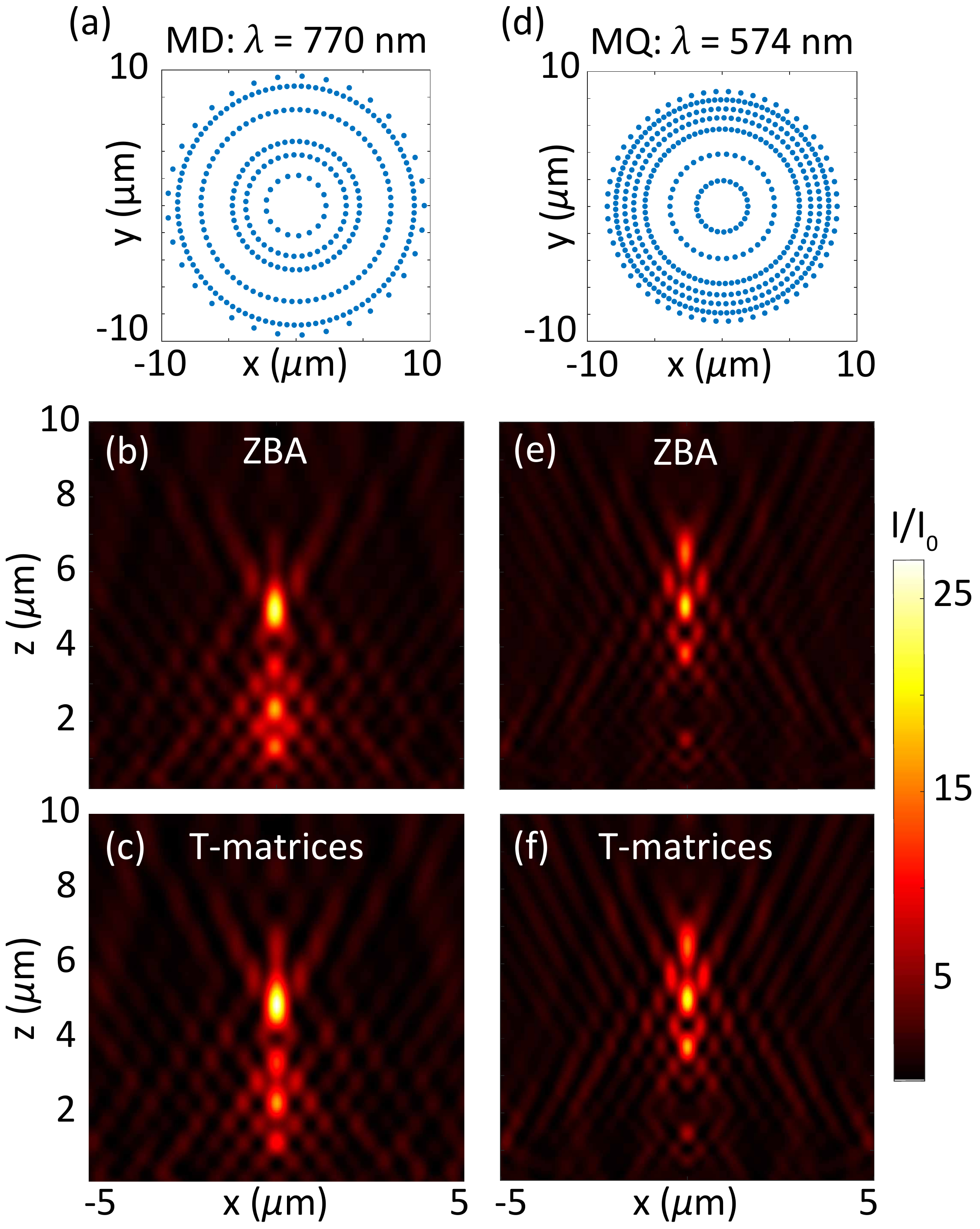}
\caption{Optimized silicon nanosphere structures for light focusing (metalenses), and their normalized intensity profiles at wavelengths of the MD resonance (a-c), and MQ resonance (d-f). (a) Scheme of particles distribution in the optimized design. (b,c) Intensity profiles of the design, calculated in ZBA (b) and by T-matrix method (c). (d-f) The same at wavelength of the MQ resonance (574 nm). The structures provide light focusing at the target position of 5 $\mu$m.}
\label{fig:optimization}
\end{figure}

Simple Evolutionary Multi-Objective Optimizer algorithm chosen by us performs a multi-criteria optimization~\cite{Kalyanmoy2002}. We have two criteria: a minimum distance between the focus of a given lens to the desired focus position (we call it a focus mismatch), and maximum focal intensity. The state of the algorithm is characterized by a set of metalenses (a population) that form a Pareto frontier (each individual metalens is not dominated by any other). One metalens \textit{dominates} another if its focus mismatch is smaller, and intensity is greater than the corresponding quantities of another lens; in all other cases, the two metalenses are considered as being ``equally good", meaning that none of the two metalenses dominates the other. On each step, one metalens is randomly chosen from the population, and then it is mutated. A mutation is an atomic change of the individual. In our problem, the mutation can affect a single ring radius, amount of particles on one ring, or initial angle deviation for particles in the ring. Once the randomly chosen individual is mutated, a Pareto frontier is updated to consists of a new set of individuals that do not dominate over each other. The stopping criterion used in this algorithm is the following: we stop the execution after a certain number of performed steps, during which the Pareto frontier is unchanged. Finally, we pick from the population the individuals with a desired focal length, and if they exist, we return the one with the highest value of focal intensity. If the algorithm failed to find such an individual, it is restarted.

The metalens designs are optimized to provide a focal length of 5 $\mu$m. The outer radius of a metalens is limited to $10$ $\mu$m, while the numbers of particles and rings are not limited. Each nanoparticle made of crystalline silicon has a fixed diameter of $200$ nm. Another external parameter is a limitation on the lowest distance between particles. The optimization process is performed in the ZBA. We consider that the incident plane wave has its polarization along the $x$-axis and propagates along the positive direction of the $z$-axis.

Figure~\ref{fig:optimization} summarizes the optimization results for two wavelengths corresponding to the MD resonance ($\lambda_{\rm MD}=770$ nm) and MQ resonance ($\lambda_{\rm MQ}=574$ nm). 
For demonstration aim, we show the results obtained with the minimum possible center-center distance between the particles during the optimization: $D = 0.92\lambda_{\rm MD}$ and $D = 1.12\lambda_{\rm MQ}$ at wavelengths of the MD resonance [Fig.~\ref{fig:optimization}(a)] and MQ resonance [Fig.~\ref{fig:optimization}(d)], respectively. Note that in optimized structures the distance between neighboring nanoparticles can be equaled or higher than the minimal values. Both structures demonstrate focusing around the desired point 5 $\mu$m. Diameters of the optimized metalenses are smaller than the predetermined limiting value (20 $\mu$m) and equal to 19.1 $\mu$m and 17.1 $\mu$m for the MD and MQ resonance structures, respectively, while the total number of particles in the structures is $N = 320$ and $N = 469$, respectively. 

\begin{figure}
    \centering
    \includegraphics[scale=0.54]{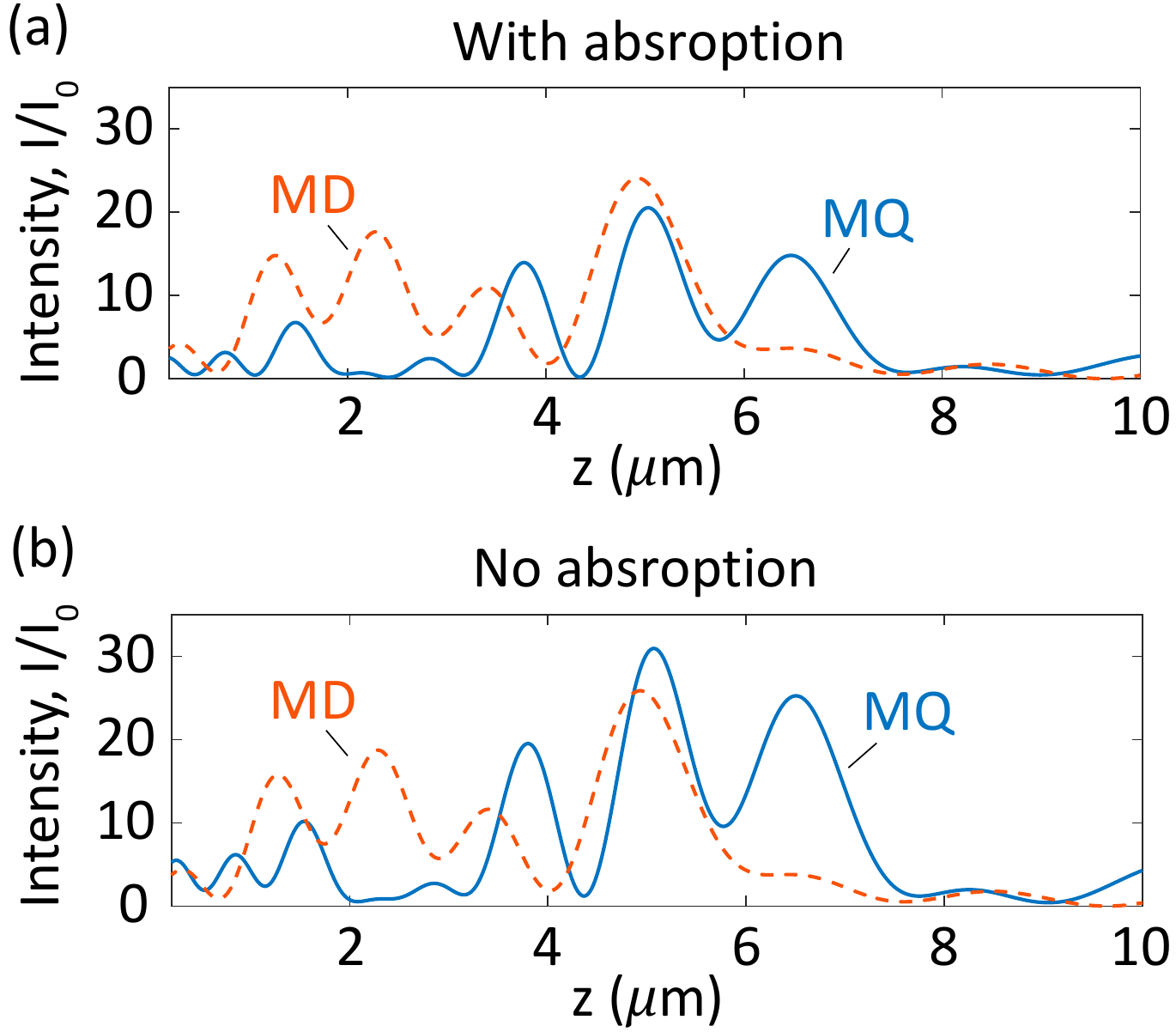}
    \caption{Normalized intensity profiles for the optimized structures in Si particles with (a), and without (b) absorption losses. The red dashed and blue lines correspond to the MD [Fig. \ref{fig:optimization}(a)] and MQ [Fig. \ref{fig:optimization}(d)] resonance structures, respectively. The intensity profiles were computed in the zero-order Born approximation by Eq.~(\ref{intensity_z_axis}) in Appendix~\ref{section:intensity}.}
    \label{fig:metalenses_I_vs_z}
\end{figure}

Despite having a smaller number of particles, the focal intensity magnitude at the MD resonance [see Fig.~\ref{fig:optimization}(b)] is higher than at the MQ resonance [see Fig.~\ref{fig:optimization}(e)]. This is affected by the absorption losses in silicon nanoparticles. Fig.~\ref{fig:metalenses_I_vs_z} presents the comparison of MD and MQ metalenses' intensity profiles with [Fig. \ref{fig:metalenses_I_vs_z}(a)] and without [Fig. \ref{fig:metalenses_I_vs_z}(b)] absorption in particles. 
For the same metalenses arranged of non-absorptive particles, the focal intensity magnitude for the MQ resonance structure is already higher than for the MD one [see Fig. \ref{fig:metalenses_I_vs_z}(b)]. Indeed, the extinction coefficient of silicon at the MQ resonance (574 nm) is almost three times higher than at the MD resonance (770 nm)~\cite{aspnes1983dielectric}.

To check the intensity profiles and optimization results, obtained the ZBA, we simulated intensity around the optimized designs [Figs.~\ref{fig:optimization}(a) and \ref{fig:optimization}(d)] by the T-matrix method~\cite{Egel2017celes} and show the results in Figs.~\ref{fig:optimization}(c) and \ref{fig:optimization}(f). A good agreement between ZBA and exact T-matrix method is obtained. Table~\ref{optimization_table} gives the focal lengths and focal intensity values of the designed metalenses obtained by the ZBA and T-matrix methods. For each structure, focal lengths are determined equally by both methods. The focus mismatch (relative to 5 $\mu$m) is relatively small: 1.6\% and 0.8\% for MD and MQ metalense, respectively. Note that the small focus mismatch of the metalenses after the optimization procedure can be reduced to zero if the radii of rings are slightly adjusted proportionally. From Table~\ref{optimization_table}, we also calculate the errors of focal intensity magnitude: 9.68\% and 7.52\% for MD and MQ resonance metalenses, respectively. Thus, we optimized the silicon nanospheres positions using an evolutionary algorithm, based on ZBA, and achieved the desired focus position with sufficiently low errors. Note that the inter-particle distance in metalens designs is higher than the particles diameter that well suitable with the laser printing method~\cite{zywietz2014laser}, which can be used for the practical fabrication of silicon nanosphere metalenses~\cite{pique2018laser}.

\begin{table}
\centering
\begin{tabular}{ |p{3.1cm}|p{1.2cm}|p{1.1cm}|p{1.2cm}|p{1.1cm}|}
 \hline
 \multirow{2}{*}{Design}& \multicolumn{2}{|c|}{ZBA}& \multicolumn{2}{|c|}{T-matrices}\\ \cline{2-5}
  & $f$ ($\mu$m)& $\mathrm{I}_f/\mathrm{I}_0$& $f$ ($\mu$m)& $\mathrm{I}_f/\mathrm{I}_0$\\
 \hline
MD: $\lambda=770$ nm, $D_{\mathrm{min}} = 0.92\lambda$& 4.92& 24.1& 4.92& 26.68\\
 \hline
MQ: $\lambda=574$ nm, $D_{\mathrm{min}} = 1.12\lambda$& 5.04& 20.53& 5.04& 22.2\\
 \hline
\end{tabular}
\caption{Summary of the actual focal length $f$ and focal intensity $\mathrm{I}_f/\mathrm{I}_0$ for optimized designs of metalenses from Figs.~\ref{fig:optimization}(a) and \ref{fig:optimization}(d). The presented quantities were calculated in ZBA and by the T-matrix method. Here the focus point is the position of the global intensity maximum.}
\label{optimization_table}
\end{table}

\section{Conclusion}\label{section:conclusion}
It was shown that the use of multipole decomposition along with the zero-order Born approximation (ZBA) allows simulation of optical properties of many-particle all-dielectric structures with good accuracy compared to exact numerical methods. The ZBA can efficiently reduce the calculation efforts, which is extremely important for optimization tasks. We analytically and numerically investigated the optical response of rings arranged by silicon nanospheres. The applicability conditions of the ZBA for calculating the focusing response of silicon nanoparticle rings at dipole and quadrupole resonances were determined. Using these conditions, we optimized structures of silicon nanospheres for a strong light focusing effect via the combination of multipole decomposition, the ZBA, and an evolutionary algorithm. It was shown that, after optimization, the non-periodical coaxial metastructure can focus  monochromatic light as metalens. The developed approach can be used for designing and optimization of metalenses and other all-dielectric structures with different functionalities. Importantly, the ZBA can be applied for designing metalenses, which can be fabricated by a laser printing technique \cite{pique2018laser} allowing to get ordered structures of well-separated spherical particles. Note that the simulation scheme based on the ZBA can be used for non-spherical nanoparticle structures as well. However, in this case, the standalone polarizabilities of a nanoparticle can be used from numerical calculations using different approaches~\cite{terekhov2017multipolar, Asadchy2014, Mun2020}.

\appendix
\section{Cartesian derivatives of Green's functions for the dipoles and quadrupoles}\label{section:green}
In this Section we provide the analytical expressions for the Cartesian derivatives of Green's functions introduced in (\ref{eq:multipole_fields}). The functions $G^{p}_{\alpha \beta}(\mathbf{r}, \mathbf{r}_0)$, $g_{\alpha}(\mathbf{r}, \mathbf{r}_0)$, $G^{Q}_{\alpha \beta}(\mathbf{r}, \mathbf{r}_0)$, and $q_{\alpha}(\mathbf{r}, \mathbf{r}_0)$ are defined by Eqs. (10), (11), (17), (18) from~\cite{Babicheva2019}, respectively.

The used notation is following: $\mathbf{r} = (x,y,z)$ is the field calculation point, $\mathbf{r}_0=(x_0,y_0,z_0)$ is the source position, $l = |\mathbf{r}-\mathbf{r}_0|$ is the point-source distance, $\mathbf{n} = (\mathbf{r}-\mathbf{r}_0)/l$ is the unit vector, $\delta_{\alpha \beta}$ is the Kronecker symbol. Greek letters $\alpha, \beta, \gamma$ denote the Cartesian coordinates $x,y,z$. 
\begin{multline}
\label{eq:d_Gpp}
\frac{\partial}{\partial \gamma} G^{p}_{\alpha \beta}(\mathbf{r}, \mathbf{r}_0) = \frac{e^{\mathrm{i} k_S l}}{4 \pi l} \times \\ \left\{ k_S \left(\mathrm{i} - \dfrac{2}{k_S l}-\dfrac{3\mathrm{i}}{k_S^2 l^2} + \dfrac{3}{k_S^3 l^3}\right)\delta_{\alpha \beta} n_{\gamma} \right. \\ \left. +k_S\left(-\mathrm{i} + \dfrac{4}{k_S l}+\dfrac{9\mathrm{i}}{k_S^2 l^2} - \dfrac{9}{k_S^3 l^3}\right)n_{\alpha}n_{\beta}n_{\gamma}
\right. \\ \left. +\left(-1 - \dfrac{3\mathrm{i}}{k_S l} + \dfrac{3}{k_S^2 l^2}\right)\left(\frac{\partial n_{\alpha}}{\partial \gamma}n_{\beta}+ n_{\alpha}\frac{\partial n_{\beta}}{\partial\gamma}\right)\right\},   
\end{multline}
\begin{eqnarray}
\frac{\partial }{\partial \gamma} g_{\alpha}(\mathbf{r}, \mathbf{r}_0) &=& \frac{e^{\mathrm{i} k_S l}}{4 \pi} \left\{ \left(- \dfrac{k_S^2}{ l}-\dfrac{2\mathrm{i}k_S}{l^2} + \dfrac{2}{ l^3}\right) n_{\alpha}n_{\gamma}  \right. \nonumber\\ &&\left. + \left(\dfrac{\mathrm{i}k_S }{l} - \dfrac{1}{l^2}\right) \frac{\partial n_{\alpha}}{\partial \gamma}\right\},    
\end{eqnarray}
\begin{multline}
\frac{\partial }{\partial \gamma} G^{Q}_{\alpha \beta}(\mathbf{r}, \mathbf{r}_0) = \frac{\mathrm{i} k_S e^{\mathrm{i} k_S l}}{24 \pi l} \times \\ \left\{k_S \left(-\mathrm{i} + \dfrac{4}{k_S l}+\dfrac{12\mathrm{i}}{k_S^2 l^2} - \dfrac{24}{k_S^3 l^3}-\dfrac{24\mathrm{i}}{k_S^4 l^4}\right)\delta_{\alpha \beta}n_{\gamma} \right.\\ +k_S\left(\mathrm{i} - \dfrac{7}{k_S l}-\dfrac{27\mathrm{i}}{k_S^2 l^2} + \dfrac{60}{k_S^3 l^3}+\dfrac{60\mathrm{i}}{k_S^4 l^4}\right)n_{\alpha}n_{\beta}n_{\gamma}
\\ \left. +\left(1 + \dfrac{6\mathrm{i}}{k_S l} - \dfrac{15}{k_S^2 l^2}-\dfrac{15\mathrm{i}}{k_S^3 l^3}\right)\left(\frac{\partial n_{\alpha}}{\partial \gamma}n_{\beta}+ n_{\alpha}\frac{\partial n_{\beta}}{\partial \gamma}\right)\right\},
\end{multline}
\begin{eqnarray}
\label{eq:d_GQM}
\frac{\partial }{\partial \gamma} q_{\alpha}(\mathbf{r}, \mathbf{r}_0) &=& \frac{k_S^2 e^{\mathrm{i} k_S l}}{24 \pi l}
\left\{ \left(1 + \dfrac{3\mathrm{i}}{k_S l} - \dfrac{3}{k_S^2 l^2}\right)\frac{\partial n_{\alpha}}{\partial \gamma} \right. \\ &&\left. +  k_S \left(\mathrm{i}- \dfrac{4}{k_S l}-\dfrac{9\mathrm{i}}{k_S^2 l^2} + \dfrac{9}{k_S^3 l^3}\right) n_{\alpha}n_{\gamma}\right\} \nonumber,
\end{eqnarray}
where derivative of the unit vector component:
\begin{gather}
    \frac{\partial n_{\alpha}}{\partial \beta} = \frac{\delta_{\alpha \beta} - n_{\alpha}n_{\beta}}{l}.
\end{gather}

\section{Intensity on the ring axis in the ZBA}\label{section:intensity}
Using Eqs. (\ref{el_dip_loc})-(\ref{mag_quad_loc}), (\ref{total_el_field})-(\ref{total_mag_field}) from the main text and Eqs. (10)-(11) from Ref.~\cite{Babicheva2019}, we derive expression for the normalized intensity along the $z$-axis:
\begin{equation}
\label{intensity_z_axis}
\mathrm{I}(z)/\mathrm{I}_0 = \tilde{\mathrm{I}}_E(z) + \tilde{\mathrm{I}}_H(z),   
\end{equation}
where the normalized electric field and magnetic field contributions in the zero-order Born approximation are following: 
\begin{eqnarray*}
    \tilde{\mathrm{I}}_E(z) &=&\frac{1}{2}\left| e^{\mathrm{i}k_S z} + N \frac{e^{\mathrm{i}k_S l}}{4 \pi l} \left\{\frac{\alpha_p}{\varepsilon_0\varepsilon_S}\left[A(l) + B(l) \frac{R^2}{2l^2}\right]\right.\right.\nonumber\\
    &&+ \alpha_m C(l) \frac{z}{l} +  \frac{\alpha_Q}{12\varepsilon_0\varepsilon_S}\frac{z}{l}\left[ D(l) + F(l)\frac{R^2}{l^2} \right]  \nonumber \\ 
    && \left.\left. + \frac{\alpha_M}{4}G(l)\frac{R^2/2 - z^2}{l^2}\right\} \right|^2, \\
\tilde{\mathrm{I}}_H(z) &=& \frac{1}{2}\left| e^{\mathrm{i}k_S z} +   N \frac{e^{\mathrm{i}k_S l}}{4 \pi l} \left\{\alpha_m \left[A(l) + B(l) \frac{R^2}{2l^2}\right]\right.\right. \nonumber\\
    &&+ \frac{\alpha_p}{\varepsilon_0\varepsilon_S} C(l) \frac{z}{l} +    \frac{\alpha_M}{4}\frac{z}{l}\left[D(l) + F(l)\frac{R^2}{l^2} \right]\nonumber\\
   && \left.\left. +  \frac{\alpha_Q}{12\varepsilon_0\varepsilon_S}G(l)\frac{R^2/2 - z^2}{l^2} \right\}\right|^2,
\end{eqnarray*}
respectively. Here $R$ is the ring radius, $l = \sqrt{R^2 + z^2}$, and:
\begin{align*}
&A(l) = k_S^2 + \frac{\mathrm{i}k_S}{l} - \frac{1}{l^2}, \quad
B(l) = -k_S^2 - \frac{3\mathrm{i}k_S}{l} + \frac{3\mathrm{i}}{l^2}, \\
&C(l) = k_S^2 + \frac{\mathrm{i}k_S}{l}, \quad
D(l) = k_S^4 + \frac{3\mathrm{i}k_S^3}{l} - \frac{6 k_S^2}{l^2} - \frac{6\mathrm{i}k_S}{l^3}, \\
&F(l) = -k_S^4 - \frac{6\mathrm{i}k_S^3}{l} + \frac{15 k_S^2}{l^2} + \frac{15\mathrm{i}k_S}{l^3}, \\
&G(l) = -k_S^4 - \frac{3\mathrm{i}k_S^3}{l} + \frac{3 k_S^2}{l^2}.
\end{align*}

\begin{acknowledgments}
The authors would like to thank M.I. Petrov, K. Frizyuk, A.V. Prokhorov for fruitful discussions, and D. Sedov for technical assistance. This work is supported by the Russian Foundation of Basic Research (RFBR) and Deutsche Forschungsgemeinschaft (DFG, German Research Foundation), Grant No. 20-52-12062, and under Germany’s Excellence Strategy within the Cluster of Excellence PhoenixD (EXC 2122, Project ID 390833453) and the Cluster of Excellence QuantumFrontiers (EXC 2123, Project ID 390837967). K.V.B. acknowledges the financial support from the German Academic Exchange Service (DAAD), and D.K. acknowledges the support from the Foundation for the Advancement of Theoretical Physics and Mathematics “BASIS”.
\end{acknowledgments}

\bibliography{arxiv_2}

\end{document}